# ESO's new generation of Exposure Time Calculators

Henri M.J. Boffin[*], Jakob Vinther,
Gurvan Bazin, David Huerta, Yves Jung, Lars K. Lundin, Malgorzata Stellert
European Southern Observatory, Karl-Schwarzschild-str. 2, 85748 Garching, Germany

## ABSTRACT

Users of astronomical observatories rely on Exposure Time Calculators (ETC) to prepare their proposals and then their observations. The ETC is therefore a crucial element in an observatory's data workflow and in particular is key to optimise the use of telescope times. This is also true for the La Silla Paranal Observatory and ESO has therefore embarked in a project to modernise its ETC, based on a python back-end and an Angular-based front-end, while also providing a programmatic interface. This ETC 2.0 has now been implemented for all the new Paranal and La Silla instruments (CRIRES, ERIS, HARPS/NIRPS, and 4MOST) and work is ongoing to implement it for MOONS. All the current ESO La Silla and Paranal instruments will also be migrated progressively, and the first one has been FORS2. The new ETC2 is based on the Instrument Packages, which should allow in the future a smooth interaction with the Phase 1 and Phase 2 observation preparation tools. Moreover, the ETC 2.0 framework has recently been upgraded and makes now use of the NgRx/Store technology in the front-end.

## 1. EXPOSURE TIME CALCULATORS

For observations at the ESO La Silla Paranal Observatory, astronomers have to apply for telescope time during Phase 1 (or p1), filling in a proposal (using the p1 tool). If the latter is successful, and they want to have these observations carried out in Service (or queue) mode [1], they are requested to submit, during Phase 2 (or p2), the associated Observing Blocks (OBs), including the requested exposure time for each target. It this therefore a requirement for the European Southern Observatory (ESO) to provide software tools for Phase 1 and 2 [2, 3]. Among these tools are the instrument exposure time calculators (ETC), which exist for each main mode of all the La Silla and Paranal instruments [4]. The ESO ETCs are web-based tools that allow users to compute the expected signal-to-noise ratio for a given observation. A presentation of the original ESO ETC is provided in [5].

Since a few years, all the ESO software tools forming the Very Large Telescope (VLT) Data flow system are being upgraded (see, e.g., [6, 7]). This is also the case of the ESO ETCs and a discussion of the major reasons for an upgrade as well as a presentation of the first implementation of this new ETC 2.0 is given in [8]. Since then, more instruments have been implemented in this new framework, and the ETC 2.0 itself underwent some major changes. We provide here an overview of what has changed since our last report.

## 2. ETC 2.0 GENERAL PRINCIPLES

One of the ultimate goals of the ETC is to allow for a smooth user experience, particularly during Phases 1 and 2, while also easing software development and maintenance. Currently, during the Phase 1 proposal preparation, astronomers need to provide information about their observations. Similar, although slightly more specific, information needs then to be provided during Phase 2, using a different tool. During both Phase 1 and Phase 2, the astronomers will likely use the ETC to estimate the time required to achieve their scientific goal. The ETC is a different tool and information will need to be copied and pasted from one tool to the other. At the Observatory, ESO astronomers may need to check the exposure times, and thus repeat the same procedure on their own. It is obvious that this process could be streamlined with current technology and this is one of the goals of the ETC 2.0 project.

---

[*] hboffin@eso.org

ETC 2.0 is a web application, where each instrument has its own interface. The front-end is implemented using Google's Angular [9], a framework for building dynamic single-page client applications using HTML and TypeScript (typed JavaScript). To allow for the different instrument ETCs to work independently, while having persistent settings, we make use of the NgRx/Store technology (see below). A screenshot of how the ETC 2.0 looks like is shown in Figure 1. The back-end is a collection of Python modules and instrument characteristic calibration data, organised in a dockerized Django web framework. In addition to the ETC calculation engines, the back-end provides services called by the front-end to dynamically configure the elements in the input forms, e.g., seeing and image quality, default extraction apertures, calculation of wavelength-shifts due to target redshift or radial velocity as well as time dependent barycentric corrections, interfacing with the SIMBAD target resolver to obtain target-specific properties.

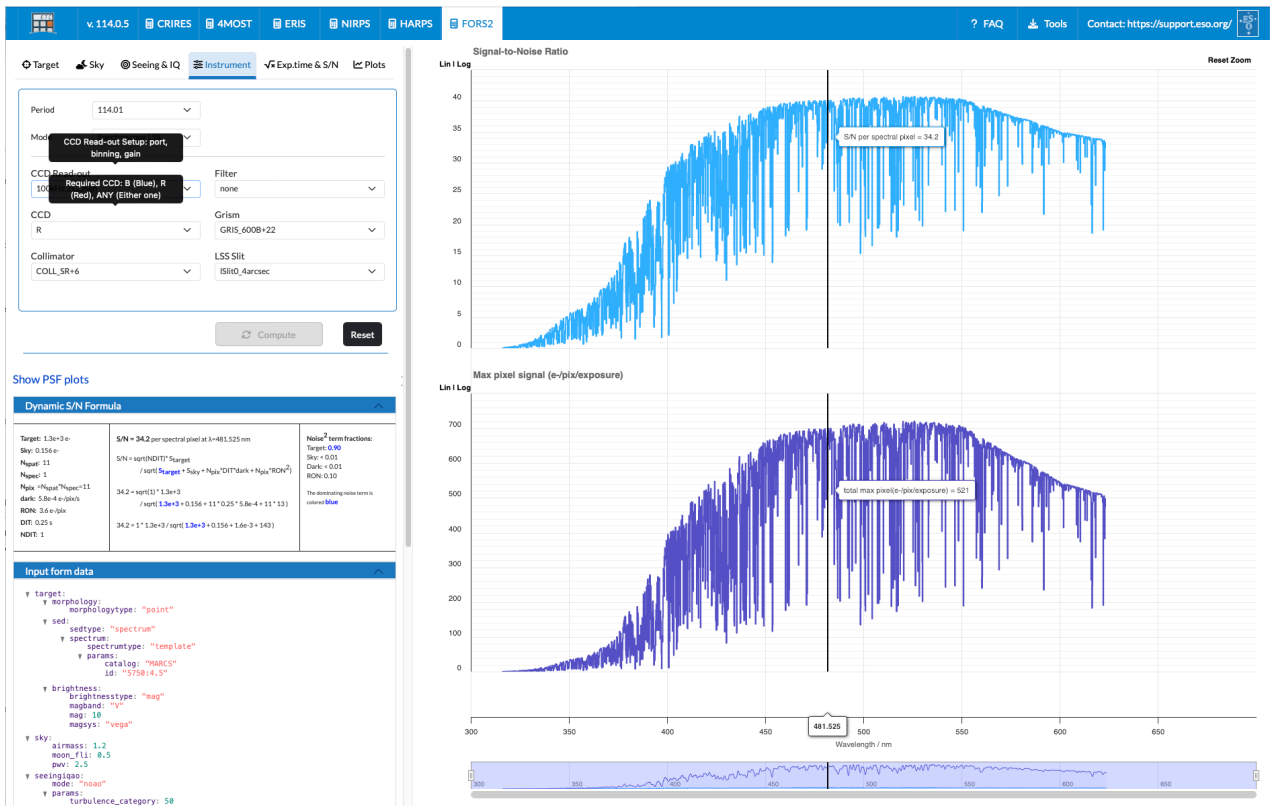

Figure 1: Screenshot of the ETC 2.0 for the FORS2 instrument.

Figure 1 highlights several of the features of the ETC 2.0:

1. The upper bar shows the various available instruments, with the current one, highlighted. One can smoothly go from one instrument to the other, and the system will remember the previous settings.

2. A set of tabs allows to define the required values for parameters related to the target itself, the sky conditions (airmass, Moon), the seeing or image quality, the exposure time or the desired signal-to-noise ratio, as well as those required specifically by the instrument (e.g., filter, grism, slit width).

3. It is also possible to choose among several plots, made using Highcharts®. Moving the cursor over the curve will provide the user with the respective value at that wavelength. Moreover, a dynamic formula can be enabled, which updates with the values corresponding to the cursor position.

Input form data

▼ target:
    ▼ morphology:
            morphologytype: "point"
    ▼ sed:
            sedtype: "spectrum"
        ▼ spectrum:
                spectrumtype: "template"
            ▼ params:
                    catalog: "MARCS"
                    id: "5750:4.5"
    ▼ brightness:
            brightnesstype: "mag"
            magband: "V"
            mag: 10
            magsys: "vega"
▼ sky:
        airmass: 1.2
        moon_fli: 0.5
        pwv: 2.5
▼ seeingiqao:
        mode: "noao"
    ▼ params:
            turbulence_category: 50
        aperturepix: 21
▼ instrument:
        slit: 0.2
        settingkey: "J1232"
        polarimetry: "free"
    ▼ order:
            0: 50
            1: 49
            2: 48
            3: 47
            4: 46
            5: 45
            6: 44
            7: 43
▼ timesnr:
        DET1.NDIT: 1
        DET1.DIT: 10
▼ output:
    ▼ throughput:
            atmosphere: false
            telescope: false
            instrument: false
            blaze: false
            enslittedenergy: false
            detector: false
            totalinclsky: false
    ▼ snr:
            snr: true
            noise_components: true
    ▼ sed:
            target: false
            sky: false
    ▼ signals:
            obstarget: false
            obssky: false
            obstotal: false
    ▼ maxsignals:
            maxpixeltarget: false
            maxpixelsky: false
            maxpixeltotal: false
    ▼ dispersion:
            dispersion: false
    ▼ psf:
            psf: false

Show          Download          Upload          Copy to clipboard

Figure 2: The ETC 2.0 uses the JSON format to exchange data with the back-end. The JSON representation of the input form can be downloaded, uploaded, or copied to the clipboard with a single action.

Furthermore, as shown in Fig. 2, there is also the possibility to show, download, upload, or copy to the clipboard the JSON data containing the input data that were set in the above-mentioned tabs. All data in the interface between front-end and back-end is indeed in JSON format.

There are several advantages to this feature. First, it is now easy to repeat exactly a given calculation, thereby allowing to share this seamlessly with colleagues. It also means that in the future, one could consider that users will simply attach their JSON file to their proposal, to show how they used the ETC to estimate the exposure times for their observations. This would then simplify the feasibility of the proposals at the Observatory. Second, users can first try the ETC online, using the web interface, then download the associated JSON file and use the provided Python script to run the ETC on the command line, using the available Application Programming Interface (API). This allows in particular running the ETC over an extended set of parameters or conditions. Finally, this also means that the communication between the ETC and the other ESO observation preparation tools, such as Phase 1 and Phase 2, is now made possible.

This latter aspect is consolidated by the fact that ETC 2.0 is now using the dedicated Instrument Packages (IPs), also available via an API [10]. The core of an IP is formed by the templates to be used for the build-up of Observing Blocks, which are the fundamental scheduling units of ESO's observatories science operations. Using a subset of the acquisition and science templates of a given instrument, one can map all the parameters needed by the ETC (such as filter, grism, CCD read-out, etc.) onto specific keywords. As shown in Fig. 1, this also means that the ETC can use the correct range of values for these parameters as well as the same mini-helps as provided in the IP. This is also done in the p2 tool. It is thus possible to envisage a not-too-distant future in which the p1 and p2 tools are directly connected with the ETC and information is smoothly transferred from one to the other. A prototype to do so between the p2 and ETC tools was already successfully made, but turning this into production may require more time.

**Additional features**

The ETC 2.0 is currently available for the most recent La Silla Paranal instruments, namely, 4MOST, CRIRES, ERIS, HARPS, and NIRPS. Soon, the ETC for MOONS will be available. The first of the older Paranal instruments, FORS2, was also recently migrated – this is the first instrument that actually uses the IP. Over the coming months, it is expected that the ETC for UVES and MUSE will also be migrated to the ETC 2.0 framework. Other Paranal instruments will follow in the next couple of years.

It is also worth mentioning some other aspects of the current ETC 2.0:

- There is a large choice of magnitude bands to define the target, using either Vega or AB magnitudes.
- There is an increased choice of templates, and we welcome suggestions for others. It is essential, however, that any template database contains spectra covering a large wavelength range and with high enough spectral resolution. Otherwise, such templates can only be used for a specific instrument.
- The ETC 2.0 uses the definition of the Paranal Turbulence Categories to define the seeing. This will therefore depend on the given instrument. Once the seeing is defined, the image quality can be computed based on the wavelength of the observations and the airmass. The ETC 2.0 then automatically adjust the extraction aperture used to compute the signal-to-noise ratio.

## 3. STATE MANAGEMENT WITH NGRX/STORE

As mentioned above, the ETC2.0 web application front-end is implemented in Angular, and much effort has been made to implement state management with NgRx/Store. As this is interesting and important aspect, we present this in more details in this section.

An Angular component consists of a TypeScript class, an HTML template, and a CSS style sheet. The TypeScript class defines the logic and holds the components' own state, while interacting with the HTML template rendered in the browser. Angular components communicate and exchange data directly or using shared services, and as such, the global application state is normally distributed across multiple components and services. In this context, the **State** is a snapshot of configurations that an application can be in at any point in time, including selected items in a form, user inputs, the state of checkboxes, components visibility, etc.

As a web application grows and evolves with additional dynamic features and external dependencies, this architecture can make the codebase complex and hard to develop and maintain. One solution is to introduce a centralized state management architecture pattern such as the NgRx/Store framework. It provides a more predictable state management pattern via the Store as the single source of truth.

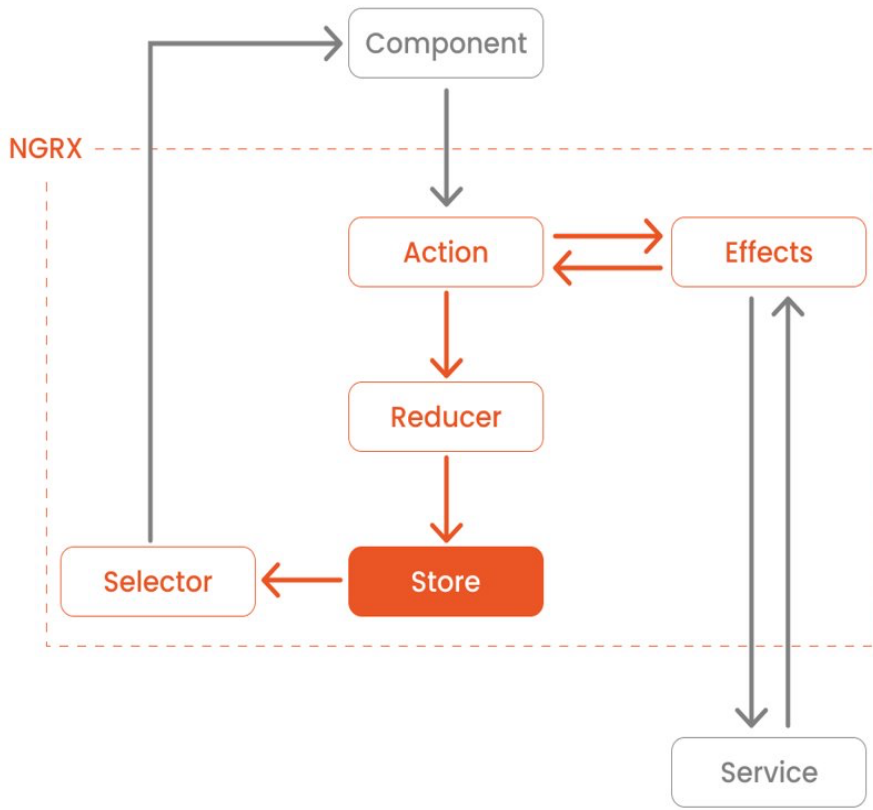

Figure 3: The main concepts in NgRx/Store. Figure reprinted with permission from [11].

The main concepts in NgRx/Store are (see Fig. 3): the State, Store, Actions, Reducers, Selectors, and Effects. **The Store** acts as a centralised container for the entire application state, maintaining a single, immutable source of truth that cannot be directly modified. Instead, when changes occur – triggered by actions dispatched to the Store – new state values are produced by reducers. These reducers are pure functions that take the current state and the action as inputs, and output a new State object with the required changes, while the existing state remains unchanged. This immutability is crucial because it helps prevent unintended side effects and makes the state changes predictable and traceable. **Actions** are events in the application, such as user interactions or network requests. Actions are dispatched by components to communicate the type of change to the Store and include as payload the information needed to update it. **Reducers** handle state changes in response to dispatched Actions, using pure functions to update the Store, returning new state objects based on previous states and the received Action, without any side effects. **Selectors** extract subsets of the state from the Store and encapsulate state query logic, which can optimize performance by caching unchanged data, preventing unnecessary recalculations of derived data. **Effects** are responsible for handling side effects of Actions, for example interactions with external resources such as HTTP calls to the ETC 2.0 back-end, or other web services. Effects can dispatch new actions back to the Store. **Components** are simpler, more linear and predictable than in plain Angular without the NgRx/Store.

**NgRx/Store in ETC 2.0**

In the ETC 2.0 web application context, we realised that the state management was getting increasingly more cumbersome in Angular without a state management system. In particular, this became clear during developing the ETC 2.0 for the HARPS and NIRPS instruments, which are mounted on the same telescope, observing the same target with the same exposure time, but in different wavelength regimes. The ETCs for these instruments were required to support mutual coupling; one instrument being the *driver* and the other the *passenger,* some of the *c*hanges made in one of them had to be propagated to the other.

It was thus necessary to introduce a state management system to support some of the requested features mentioned elsewhere in this paper, including:

- State preservation in individual ETCs;
- Capability to save and preset the form configurations;
- Interface with other applications (e.g., the p2 tool).

It was decided to introduce the NgRx/Store state management design. The implementation has required a substantial code base refactoring, but in terms of software development and maintenance it has proven to be worth it:

- Developing and implementing new ETCs can be done faster and more consistently;
- Easier maintenance;
- Better debugging tools ("time travel" inspection between states).

However, NgRx/Store has downsides to be considered as well:

- Complexity and Overhead: it is overkill for small or simple applications;
- Steep learning Curve: Redux-style state management is not trivial;
- Verbose: "boiler-plate" repetitive code is needed by the framework/workflow, without being specific to requirements.